%% file: 2019-VPT-LemmaGen-PostProc.tex
\documentclass[submission,copyright,creativecommons]{eptcs}
\providecommand{\event}{          
    Verification and Program Transformation (VPT 2019)
}
\usepackage{breakurl}             
\usepackage[strings]{underscore}  
\usepackage{color}

\title{Lemma Generation for Horn Clause Satisfiability:\\
A Preliminary Study}   

\author{
Emanuele De Angelis
\institute{DEC, University ``G.~d'Annunzio" of Chieti-Pescara\\[2pt]
    Viale Pindaro 42, 65127 Pescara, Italy
}
\email{emanuele.deangelis@unich.it}
\and
Fabio Fioravanti
\institute{DEC, University ``G.~d'Annunzio" of Chieti-Pescara\\[2pt]
    Viale Pindaro 42, 65127 Pescara, Italy
}
\email{fabio.fioravanti@unich.it}
\and
Alberto Pettorossi
\institute{
    University of Roma Tor Vergata\\[2pt]
    Via del Politecnico 1, 00133 Roma, Italy
}
\email{pettorossi@info.uniroma2.it}
\and
Maurizio Proietti
\institute{
    CNR-IASI\\[2pt]
    Via dei Taurini 19, 00185 Roma, Italy
}
\email{maurizio.proietti@iasi.cnr.it}
}

\def\titlerunning{Lemma Generation for Horn Clause Satisfiability}

\def\authorrunning{E.~De Angelis et al.}

\begin{document}
\maketitle

\begin{abstract}
\input{sections/0_abstr.tex}
\end{abstract}

\section{Introduction}
\input{sections/1_intro.tex}

\section{Horn Clause Satisfiability for Program Verification}
\input{sections/2_chcsat.tex}

\section{Introducing Difference Predicates}
\input{sections/3_sumlist.tex}

\section{Introducing Auxiliary Queries}
\input{sections/35_rotate.tex}

\section{Concluding Remarks}
\input{sections/4_conclremks.tex}

\section{Acknowledgments}
\input{sections/5_acknowledgement.tex}

\input{sections/6_Biblio.tex}

\end{document}

%% file: sections/0_abstr.tex
\begin{abstract}
It is known that the verification of imperative, functional, and logic
programs can be 
reduced to the satisfiability of constrained Horn clauses (CHCs), and this
satisfiability check can be performed by using 
CHC solvers, such as Eldarica and~Z3.
These solvers perform well when they act on simple constraint theories,
such as Linear Integer Arithmetic and the theory of Booleans, 
but their efficacy is very much reduced when the clauses
refer to constraints on inductively defined structures, such as lists or trees. 
Recently, we have presented a transformation technique for eliminating 
those inductively defined data structures, and hence avoiding the need for incorporating
induction principles into CHC solvers.
However, this technique may fail when the transformation
requires the use of lemmata whose generation needs ingenuity.
In this paper we show, through an example, how during the process 
of transforming CHCs for eliminating inductively defined structures
one can introduce suitable predicates, called difference predicates,
whose definitions correspond to the lemmata to be introduced.
Through a second example, we show that, whenever difference predicates cannot
be introduced, we can introduce, instead,
auxiliary queries which also correspond 
to lemmata, and the proof of these lemmata can be done by showing the
satisfiability of those queries.
\end{abstract}

%% file: sections/1_intro.tex
\label{intro}

In recent years, it has been shown that the verification of program 
properties can be performed by proving the satisfiability of sets of   
{\em{constrained Horn clauses}} (CHCs). Since  a general decision 
procedure for proving satisfiability of CHCs does not exist, the best one can do is
to propose heuristics, and indeed various 
heuristics for  proving satisfiability  
have been proposed in the literature. Among them we recall:
 (i)~{\rm Counterexample Guided 
Abstraction Refinement} (CEGAR)~\cite{Cl&00}, (ii)~{\rm Craig interpolation}~\cite{McM03}, and 
(iii)~{\rm Property Directed 
Reachability} (PDR)~\cite{Bra11,HoB12}.
Moreover, a variety of tools for satisfiability proofs, 
called {\em{CHC solvers}}, 
has been made available to the scientific community. Let us mention:
Eldarica~\cite{Ho&12}, HSF~\cite{Gr&12}, RAHFT~\cite{Ka&16}, 
VeriMAP~\cite{De&14b}, and Z3~\cite{DeB08}. Most of those tools work well on simple
constraint theories, such as the 
theory of Linear Integer Arithmetic ($\mathrm{LIA}$) and the theory 
of Booleans~($\mathrm{Bool}$).

Unfortunately, when the properties to be verified refer to programs 
that act on inductively 
defined data structures, such as lists or trees, then 
the satisfiability proofs via CHC solvers become much harder, 
or even impossible, 
because those solvers do not usually incorporate  
induction principles relative to the data structures in use.

To avoid this difficulty, two approaches have recently been suggested. 
The first one consists in the
incorporation into the CHC solvers of appropriate 
induction principles~\cite{ReK15,Un&17}.
The second one consists in transforming the given set of CHCs 
into a new set where inductively defined data structures are removed, and whose satisfiability
implies the satisfiability of the original clauses~\cite{De&18a,MoF17}.

In this paper we will follow this second approach and, 
in particular, we will consider the 
{\em Elimination Algorithm} presented in a previous work of ours~\cite{De&18a},
which implements a transformation strategy for removing inductively defined data
structures and is 
based on the familiar fold/unfold rules~\cite{EtG96,TaS84}.
Thus, if the clauses derived by the Elimination Algorithm have all 
their constraints 
in the {\rm{LIA}} or {\rm{Bool}} theory, then
there is no need to modify the CHC solvers for performing 
the required satisfiability proofs.

Similarly to the case of proof techniques that use induction, 
the success of the Elimination Algorithm may depend on the
discovery of suitable auxiliary lemmata to be used during transformation. 
The contribution of this paper is a novel technique
for generating those lemmata, thereby providing a way of
overcoming some difficulties which have been reported
in the literature and, in particular, the need of dealing
with formulas with second order variables and second order unification,
when performing inductive proofs~\cite{Bun01}.

This novel technique
is presented through two examples. In the first example, we show that 
the introduction of suitable 
predicates, which we call {\it difference predicates}, may allow us 
to perform the desired transformations. A difference predicate
expresses the relation between the values computed by two different functions
(hence its name), and its definition
corresponds to the statement of a lemma which should be proved
if one were to show the properties of interest by structural induction.
In this example it is demonstrated that, by extending the Elimination Algorithm
with the introduction of difference predicates, one can remove inductively 
defined data structures, thus allowing the completion of the desired proof
in many cases where the plain Elimination Algorithm would not terminate. 

In the second example, we show how to introduce, during 
CHC transformation, some implications which correspond to suitable 
lemmata. Those implications are then proved by showing the satisfiability of
auxiliary queries (that is, clauses whose head is {\tt false}), which are derived from the implications.

\smallskip
The paper is structured as follows. In Sections~\ref{chcsat} and~\ref{diffpred},
we present the verification of
a property of a functional program that acts on lists of integers, 
by first (i)~deriving 
by transformation, introducing a suitable difference predicate,
a set of CHCs on LIA constraints only (that is, constraints on lists
will no longer be present), and then
(ii)~proving the satisfiability of the derived CHCs by using the solver~Z3
acting on LIA constraints only.
Note that neither~Z3 nor~Eldarica are able to check satisfiability of 
the clauses which are obtained by the direct translation into CHCs
of the functional program and the property, before the 
transformation of Step~(i).
In Section~\ref{rotate}, we present a second example of our 
verification technique where, during Step~(i), we introduce,
instead of difference predicates,  suitable auxiliary queries.
Finally, in Section~\ref{conclrems},  we comment on the soundness and the 
mechanization of our verification technique.

%% file: sections/2_chcsat.tex
\label{chcsat}
Let us consider the following functional program {\it InsertionSort}, written in the OCaml syntax~\cite{Le&17}:

\begin{verbatim}
type list = Nil | Cons of int * list
let rec ins i l =
  match l with
  | Nil -> Cons(i,Nil)
  | Cons(x,xs) -> if i<=x then Cons(i,Cons(x,xs)) else Cons(x,ins i xs)
\end{verbatim}
\vspace*{-6.5mm}
\begin{verbatim}
let rec insertionSort l =
  match l with
  | Nil -> Nil
  | Cons(x,xs) -> ins x (insertionSort xs)
let rec sumlist l =
  match l with
  | Nil -> 0
  | Cons(x,xs) -> x + sumlist xs
\end{verbatim}

\noindent
In this program: (i)~the  {\tt insertionSort} function sorts a list of integers, in ascending order, according to the
familiar insertion sort algorithm, and (ii)~the {\tt sumlist} function
computes, given a list of integers, the sum of all integers in that list.

Let us suppose that for the program {\it InsertionSort} we want to prove the 
following Property~$\mathit{Sum}$  stating that the sum of 
the elements of a list~$\mathtt{l}$
is equal to the sum of the elements of the sorted list 
$\mathtt{insertionSort~l}$. 
Thus, in formulas, we want to prove that:

\smallskip

$\mathtt{\forall\, l.~~ sumlist~l = sumlist~(insertionSort~l)}$ \hfill 
\makebox[32mm][l]{$($Property~$\mathit{Sum})$}

\smallskip

\noindent
If we want to make a  proof of Property~${\mathit{Sum}}$
by induction on the structure of the list~$\mathtt{l}$, 
we have to use a lemma stating that the sum of the elements of the list 
{\tt{ins x l}} obtained
by inserting the element {\tt x} in the list {\tt l} 
is obtained by adding {\tt x} to 
the sum of the elements of {\tt l}. This lemma can be expressed 
by the following formula:
\smallskip

$\mathtt{\forall\, x, l.~~  sumlist~(ins~x~l) = x + (sumlist~l)}$ \hfill 
\makebox[32mm][l]{$($Lemma~$\mathit{L})$}

\smallskip

\noindent
The technique we present in this paper for the proof of 
Property~${\mathit{Sum}}$, avoids the explicit introduction of this lemma,
and thus the use of the induction principle on lists. 

Let us start off by considering the translation of the functional program 
${\mathit{InsertionSort}}$ 
and Property~${\mathit{Sum}}$ into a set of 
CHCs as explained in the literature~\cite{De&18a,Un&17}. 
In our example, by that translation we get the following set of clauses\,\footnote{~We use 
{Prolog-like} syntax for writing clauses, instead of the more verbose
SMT-LIB syntax. 
The predicates {\texttt =}~(equal), 
\mbox{\texttt{$\neq$}~(not-equal)},  {$\leq$}~(less-or-equal), and
{\tt >}~(greater)
denote constraints between integers.}:

\begingroup
\catcode`\@=\active
\def @{$\,\neq\,$}        

\catcode`\¢=\active
\def ¢{$\,\leq\,$}        

\begin{verbatim}
 1. false :- M@N, sumlist(L,M), insertionSort(L,SL), sumlist(SL,N).
 2. sumlist([],0).
 3. sumlist([X|Xs],M) :- M=X+N, sumlist(Xs,N).
 4. ins(I,[],[I]).
 5. ins(I,[X|Xs],[I,X|Xs]) :- I¢X.
 6. ins(I,[X|Xs],[X|Ys]) :- I>X, ins(I,Xs,Ys).
 7. insertionSort([],[]). 
 8. insertionSort([X|Xs],SL) :- insertionSort(Xs,SXs), ins(X,SXs,SL).
\end{verbatim}
\endgroup

\noindent
In these clauses, $\mathtt{sumlist(L,M)}$,
$\mathtt{insertionSort(L,SL)}$, and 
$\mathtt{ins(X,L,L1)}$ hold ~iff~ 
$\mathtt{sumlist~L}$ $=$ $\mathtt M$, $\mathtt{insertionSort~L}$ $=$ 
$\mathtt{SL}$, and 
$\mathtt{ins~X~L}$ $=$ $\mathtt{L1}$, respectively, hold in 
program ${\mathit{InsertionSort}}$.

As usual, we assume that all clauses are universally quantified in front.
Clause {\tt1}, also called a {\em query},$\!$\footnote{~In the context of Horn clauses, a query (or a goal) is a clause whose head 
is {\tt false}.}
translates Property~${\mathit{Sum}}$ as it stands  
(using the functional notation) for:

\smallskip

$\mathtt{\forall\, l,m,sl,n.~\ {sumlist}~l = m ~\wedge~ {insertionSort}~l = sl 
~\wedge~  {sumlist}~sl = n ~~\rightarrow~~ m = n } $

\smallskip

\noindent
and clauses~{\tt1}--\,{\tt8} are satisfiable iff 
Property~${\mathit{Sum}}$ holds. 
Unfortunately, state-of-the-art CHC solvers, such as Eldarica or Z3, 
fail to prove satisfiability of clauses~{\tt1}--\,{\tt8},
because those CHC solvers do not incorporate any induction principle on lists.

Moreover, starting from clauses {\tt1}--\,{\tt8}, the Elimination
Algorithm~\cite{De&18a} 
which has the objective of eliminating lists from CHCs, is not able to 
derive a set of clauses without lists, and this inability is
due to the fact that the algorithm is unable to introduce 
a predicate definition corresponding to the needed
Lemma~$L$.

%% file: sections/3_sumlist.tex
\label{diffpred}
\newcommand{\bbar}{{|\hspace*{-.4mm}|}}

In this section we show that, if we extend the Elimination Algorithm~\cite{De&18a} 
by a technique for introducing the so-called difference predicates,
we are able to
transform clauses~{\tt1}--\,{\tt8} into a set of clauses without list variables.
We will see that the 
definition of the difference predicate we will introduce exactly
corresponds to Lemma~$L$.

First, we briefly recall the Elimination Algorithm which makes use of the
well-known transformation rules:
(i)~{\em define}, (ii)~{\em fold}, (iii)~{\em unfold}, and (iv)~{\em replace} for 
CHCs~\cite{EtG96,TaS84}. The details can be found in the paper where 
the algorithm was originally
presented~\cite{De&18a}. 

We assume that the {\textit{basic types}} 
are the integers and the booleans. 
We say that a clause has basic types if all its variables have
basic types. 
In the outline of the algorithm below, {\it Cls} is the set of input clauses which 
define the predicates occurring
in the given set {\it Qs} of input queries. {\it Defs} is the set of
definition clauses which are introduced by the algorithm and used for folding. 
{\it Defs} accumulates the sets {\it NewDefs} of
definition clauses which are introduced during the various iterations of
the while-do loop.
\mbox{\it FldCls}, {\it UnfCls}, and {\it RCls} are the 
sets of clauses which are
obtained after the applications of the folding, unfolding, and replace rules, 
respectively. The Elimination Algorithm works by enforcing that all new 
predicate definitions which are introduced have arguments of basic types 
only (and thus, no list variables will occur in their arguments).

\noindent \hrulefill\nopagebreak

\noindent {\bf The Elimination Algorithm}~$\mathcal E$.\\
{\em Input}: A set $\mathit{Cls} \cup \mathit{Qs}$, where $\mathit{Cls}$ is a set of constrained Horn clauses and $\mathit{Qs}$ 
is a set of queries.
\\
{\em Output}: A set $\mathit{TransfCls}$ of clauses such that: 
(i)~$\mathit{Cls}\cup \mathit{Qs}$ is satisfiable iff $\mathit{TransfCls}$ is satisfiable, and
(ii)~every clause in $\mathit{TransfCls}$ has basic types.

\vspace*{-2mm}
\noindent \rule{2.0cm}{0.2mm}

\noindent $\mathit{Defs}:=\emptyset$;
\noindent $\mathit{InCls}:=\mathit{Qs}$;
\noindent $\mathit{TransfCls}:=\emptyset;$

\noindent
{\bf while} ~~$\mathit{InCls}\!\neq\!\emptyset$~~ {\bf do}

$\mathit{Define\mbox{-}Fold}(\mathit{Defs},\mathit{InCls},
\mathit{NewDefs},\mathit{FldCls});$

$\mathit{Unfold}(\mathit{NewDefs},\mathit{Cls},\mathit{UnfCls});$

$\mathit{Replace}(\mathit{UnfCls}, \mathit{Cls}, \mathit{RCls});$

$\mathit{Defs}:=\mathit{Defs}\cup\mathit{NewDefs};$~~
$\mathit{InCls}:=\mathit{RCls};$~~
$\mathit{TransfCls}:=\mathit{TransfCls}\cup\mathit{FldCls};$\nopagebreak

\vspace*{-2mm} 
\noindent \hrulefill

\smallskip

In order to get clauses without list variables,
we start off the transformation of clauses {\tt1}--\,{\tt8} by applying the 
{Elimination Algorithm}~$\mathcal E$ to the following sets of clauses and queries: {\it Cls} $=$ 
\{clause~{\tt2}, \ldots,
clause~{\tt8}\} and {\it Qs} $=$ \{query~{\tt1}\}.
Thus, the first step is the introduction of a new predicate 
{\tt new1} by the following clause 
(here and in what follows the numbers under the atoms of the body of the clauses 
identify the individual atoms):
\begin{verbatim}
9. new1(M,N) :- sumlist(L,M), insertionSort(L,SL), sumlist(SL,N).
                   (9.1)             (9.2)           (9.3)
\end{verbatim}

\noindent
The arguments of $\mathtt{new1}$ are the integer variables occurring in 
the body of clause~{\tt9}.
Thus, by folding query~{\tt1}, we derive a new clause without occurrences of list variables:

\begingroup
\catcode`\@=\active
\def @{$\neq$}    

\begin{verbatim}
10. false :- M@N, new1(M,N).
\end{verbatim}
\endgroup

\noindent
We proceed by eliminating lists from the body of clause~{\tt9}. 
By unfolding clause~{\tt9}, we replace the predicate calls by their 
definitions and we derive the following two clauses:


\begin{verbatim}
11. new1(0,0).
12. new1(M1,N1) :- M1=H+M, sumlist(T,M), insertionSort(T,ST), ins(H,ST,SU), 
                             (12.1)           (12.2)            (12.3)  
                           sumlist(SU,N1).        
                             (12.4)
\end{verbatim}

\noindent
Now, in order to fold clause~{\tt12} using clause~{\tt9} and 
derive a recursive definition of {\tt new1}, 
we depart from the Elimination Algorithm
and we propose a new technique that introduces a so-called {\em difference
predicate}~{\tt diff}. The definition of~{\tt diff} is based on the mismatch 
between clause~{\tt9} and clause~{\tt12}. The new technique is 
applied according to the following six steps.

\smallskip
\noindent
$\bullet$ {\it Step} 1. {\it Embed}. 
We have that the body of clause~{\tt9} is {\em embedded}
in the body of clause~{\tt12}, that is,  each distinct atom 
in the body of clause~{\tt9} is a variant of a distinct atom 
in the body of clause~{\tt12}. In particular,  
(i)~{\tt9.1} is a variant of~{\tt12.1},
(ii)~{\tt9.2} is a variant of~{\tt12.2}, and
(iii)~{\tt9.3} is a variant of~{\tt12.4}.
However, clause~{\tt12} cannot be folded using clause~{\tt9},
because the conjunction ({\tt9.1}, {\tt9.2}, {\tt9.3}) does not match
the conjunction ({\tt12.1}, {\tt12.2}, {\tt12.4})
(see arguments {\tt SL}, {\tt ST}, and {\tt SU}).
It can be shown that no further unfolding of clause~{\tt12} will generate
a clause whose body is an instance of the body of clause~{\tt9},
and indeed the Elimination Algorithm will not terminate.

\smallskip
\noindent
$\bullet$ {\it Step} 2. {\it Rename}. We rename apart clause~{\tt 9}, 
which we would like to use for folding, 
so as to have variable names that do not occur anywhere else\footnote{\,This
is the only variable renaming that we perform during Steps~1--\,6.}. We get:

\begin{verbatim}
9a. new1(Ma,Na) :- sumlist(La,Ma), insertionSort(La,SLa), sumlist(SLa,Na).                             
                     (9a.1)                (9a.2)            (9a.3)
\end{verbatim}

\noindent
$\bullet$ {\it Step} 3. {\it Match}. We match the body of clause~{\tt9a}  
against 
the body of clause~{\tt12} to be folded. We match the conjunction 
({\tt9a.1}, 
{\tt9a.2}) with 
the conjunction ({\tt12.1}, {\tt12.2}) by the renaming substitution 
$\sigma= \{{\mathtt{La/T,}}$ ${\mathtt{Ma/M,}}$ 
${\mathtt{SLa/ST}}\}$, 
but we cannot extend this matching
to the remaining atom {\tt9a.3}
because the substitution $\{{\mathtt{SLa/SU,~Na/N1}}\}$ 
is inconsistent with $\sigma$. 
By applying the substitution~$\sigma$, we get the following clause~{\tt9m}, which 
is a variant of clause~{\tt9a}:

\smallskip
\noindent
${\mathtt{9m.~~new1(M,Na)~:\mbox{\tt -}~sumlist(T,M),~insertionSort(T,ST),
   ~\bbar~sumlist(ST,Na).}}$ 
   
\noindent
${\mathtt{\hspace{38mm}(9m.1) \hspace{20mm} (9m.2)\hspace{28mm}(9m.3)}}$    

\smallskip
\noindent
This clause~{\tt 9m} is the actual
clause which we will use for folding at Step~6 below.
The marker~$\bbar$~we have placed in its body 
has no logical meaning and it is used only as a separator between
 the {\it matching conjunction} 
({\tt9m.1}, {\tt9m.2}) 
to its left and the {\it mismatching conjunction} 
\mbox{\tt9m.3} 
to its right (in general, also the {\rm mismatching conjunction} may 
consist of more than
one atom). 

Also for the clause to be folded (clause~{\tt12} in our case)
we define the matching and the mismatching conjunctions:
(i)~the {\it matching conjunction} of the clause to be folded 
is equal to the one of the clause
we will use for folding (atoms {\tt 12.1} and {\tt 12.2} in our case),
while (ii)~the {\it mismatching conjunction}  of the clause to be folded 
is the conjunction of its body atoms 
(atoms~{\tt12.3} and~{\tt12.4} in our case) that do not belong
to the {\rm matching conjunction}.

\smallskip
\noindent
$\bullet$ {\it Step}~4. {\it Introduce a Difference Predicate}. 
Now, in order to fold 
clause~{\tt12} using clause~{\tt 9m}, we need to replace some atoms of the
body of clause~{\tt12}. In particular,

\noindent
{\it Point}~(4.1): we have to remove from the body of clause {\tt12} 
its mismatching conjunction (atoms {\tt12.$\!$3} and {\tt12.$\!$4}), 
and 

\noindent
{\it Point}~(4.2): we have to add to the body
of clause {\tt12} the mismatching conjunction of clause~{\tt9m}
(atom~{\tt9m.3}). 

Indeed, if we perform the actions of Points~(4.1) and (4.2) 
(see Step~5 below), then
we can fold clause~{\tt12} by using clause~{\tt9m} (see Step~6 below).
However, before performing that folding step, 
we also add to the body of clause~{\tt12} a
new atom defining a so-called {\it difference predicate}.
This new atom expresses the relation between the output non-list 
variables\,\footnote{~The output 
variables of the 
predicates {\tt ins} and {\tt sumlist} in the atoms {\tt 12.3}, {\tt 12.4}, and 
{\tt 9m.3} 
are defined as expected, when considering the associated functional expressions 
{\tt (ins~H~ST)\,=\,SU}, {\tt (sumlist~SU)\,=\,N1}, and 
\mbox{\tt (sumlist ST)\,=\,Na}, respectively.} of the 
conjunction to be removed according to Point~(4.1) (that is, 
the integer variable~{\tt N1}, 
in our case),  and the output non-list 
variables  of the conjunction to be added according to Point~(4.2) (that
is, the integer variable~{\tt Na}, in our case). 
%
%
%
The difference 
predicate we introduce, called {\tt diff}, is defined by a clause, 
whose body {is made out of: (i)~the mismatching conjunction 
to be removed, and 
(ii)~the mismatching conjunction to be added. Thus, 
we get the following definition for {\tt diff}:}\nopagebreak


\smallskip
\noindent
${\mathtt{13.~~diff(H,Na,N1)~:\mbox{\tt -}~ins(H,ST,SU),~sumlist(SU,N1),}}
~~{\mathtt{sumlist(ST,Na).}}$\nopagebreak

${\mathtt{\hspace{40mm}(12.3) \hspace{12mm} (12.4)\hspace{18mm}(9m.3)}}$

\smallskip

\noindent
The arguments {\tt H}, {\tt Na}, and {\tt N1} of the head 
are the non-list variables 
occurring in the body 
(obviously these arguments
can be placed in any order)\,\footnote{~Note that this choice of 
the arguments of {\tt diff}
is in accordance with our goal of eliminating list variables.}. 
Clause~{\tt13} 
can be read (in functional notation) as follows:






\smallskip

 {\it{if}\/} $\mathtt{sumlist~(ins~H~ST)\,=\,N1}$ ~and~ 
$\mathtt{sumlist~ST\,=\,Na}$, 

\hangindent=14mm
{\it{then}\/} $\mathtt{diff(H,Na,N1)}$ expresses the relation between the output 
integer variables {\tt Na} and {\tt N1} (as a function of the input integer variable {\tt H}).


\smallskip
\noindent
$\bullet$ {\it Step} 5. {\it Replace}.  In the clause to be folded 
(clause~{\tt 12} in our case) we replace the mismatching conjunctions 
(atoms~{\tt12.3} and~{\tt12.4} in our case) by: 
(i)~the mismatching conjunction of the clause we will use for folding 
(atom~{\tt9m.3} in our case), and 
(ii)~the head of the clause defining the difference predicate
(clause~{\tt 13} in our case). 
By doing so, from clause~{\tt 12} we get the following clause:
 
\begin{verbatim}
12r. new1(M1,N1) :- M1=H+M, sumlist(T,M), insertionSort(T,ST), sumlist(ST,Na), 
                    diff(H,Na,N1).
\end{verbatim}

\noindent
$\bullet$ {\it Step} 6. {\it Fold}. We fold clause~{\tt12r} using clause~{\tt9m}
(this folding is possible due to Step~5) and we get:

\begin{verbatim} 
12f. new1(M1,N1) :- M1=H+M, new1(M,Na), diff(H,Na,N1).
\end{verbatim}

\noindent

Now let us briefly discuss the correctness of the above 
transformation consisting of Steps 1--\,6.

In general, we say that
a transformation from a given set $S_{1}$ of clauses to a derived set $S_{2}$ of clauses 
is {\it sound} whenever: 

\smallskip

{\it if}\/ $S_{2}$ is satisfiable, ~{\it then} $S_{1}$ is satisfiable. 
\hfill\makebox[30mm][l]{({\it Soundness})}

\smallskip
\noindent
Thus, the satisfiability of the derived set of clauses
is sufficient to guarantee the property of the 
functional program that we want to verify (Property {\it Sum}, in our 
example).

In the case of our transformation Steps 1--\,6, soundness is ensured by the fact that, at Step~5, 
we replace a given conjunction of atoms
by a new conjunction which is implied by the given one.
For instance, in our {\it InsertionSort} example, we have that:

\smallskip

\noindent
{\tt I.}~~ $\mathtt{\forall\, (ins(H,ST,SU),~sumlist(SU,N1) \rightarrow (\exists\, Na.~sumlist(ST,Na),~diff(H,Na,N1)))}$

\smallskip
\noindent
where, for any formula $\varphi$, $\mathtt{\forall (\varphi)}$
denotes the {\it universal closure} of $\varphi$, that is,
the formula $\mathtt{\forall \overline{X}.\, \varphi}$, where 
$\mathtt{\overline{X}}$ is the tuple of 
the variables occurring free in $\mathtt{\varphi}$.
Indeed, formula~{\tt I} holds in the least model of
clauses~\mbox{{\tt2}\,--\,{\tt8}}, because: (i)~for all {\tt L} and {\tt N},
$\mathtt{sumlist(L,N)}$  
defines a total functional relation from 
(the domain of) {\tt L} to (the domain of) {\tt N} (and thus,  
$\mathtt{\exists\, Na.~sumlist(ST,Na)}$ is true),
and (ii)~the predicate {\tt diff} is defined by clause~{\tt 13}.

\smallskip
It can also be shown that the above Steps 1--\,6  preserve
satisfiability of clauses in the following stronger sense: 


\smallskip
{\it{if}}\makebox[11.mm][r]{(H1)~~}$\mathtt{ins(X,S,S1)}$ and 
$\mathtt{sumlist(L,N)}$ define total
functional relations,  \hfill\makebox[30mm][l]{({\it Equisatisfiability})}

\hspace{5mm}{\it{and}\/} 

\hangindent=7.5mm
\makebox[13.3mm][r]{(H2)~~}${\mathtt{diff(H,Na,N1)}}$ is a functional 
relation from the pair {\tt (H,Na)} of integers to the integer
{\tt N1},

{\it{then}\/} the replacement of clause~{\tt12} by clauses 
{\tt12f} and~{\tt13}  produces an {\it equisatisfiable} set of clauses.

\smallskip
\noindent
Indeed, under Hypotheses (H1) and (H2), 
also the converse of implication {\tt I} holds.

\smallskip
Note that Hypothesis (H1) holds by construction, because the predicates 
$\mathtt{ins}$ and $\mathtt{sumlist}$ come from the translation of 
functional programs 
that terminate for all input values.

The detailed proofs of general results concerning the soundness and equisatisfiability 
properties of our transformations are outside the scope of the present paper.


\medskip
The clauses we have derived so far are clauses {\tt10}, {\tt11}, {\tt12f}, 
{\tt13}, together with the clauses defining the predicates
 {\tt sumlist} and {\tt ins}, that is,
clauses~{\tt2}\,--\,{\tt6}. Still clause~{\tt13},
which defines the predicate {\tt diff}, and clauses~{\tt2}\,--\,{\tt6}
have list variables and we should eliminate them
by applying the
Elimination Algorithm. If we do so by starting from
{\it Cls} $=$ \{clause~{\tt2}, \ldots, clause~{\tt6}\}
and {\it InCls} $=$ \{clause~{\tt13}\},  we derive the following final clauses 
(during this elimination there is no need of introducing any new difference predicate):

\begingroup
\catcode`\@=\active
\def @{$\neq$}    

\catcode`\¢=\active
\def ¢{$\leq$}     

\begin{verbatim}
10.  false :- M@N, new1(M,N).
11.  new1(0,0).
12f. new1(M1,N1) :- M1=H+M, new1(M,Na), diff(H,Na,N1).
14.  diff(H,0,N1) :- N1=H.
15.  diff(H,Na,N1) :- H¢X, Na=X+N2, N1=H+Na, new2(N2).
16.  diff(H,Na,N1) :- H>X, Na=X+N2, N1=X+N3, diff(H,N2,N3).
17.  new2(0).
18.  new2(N) :- N=X+N1, new2(N1).
\end{verbatim}
\endgroup

\noindent
This final set of clauses
 has no list arguments and all the constraints belong to the LIA theory.
The CHC solver Z3\,\footnote{~By first translating CHCs from Prolog syntax to SMT-LIB syntax
and then using the command:
`\texttt{z3\_4.8.4 -smt2 sumlist.transf.smt fp.engine=spacer dump\_models=true}'.
}  proves that this set of clauses is satisfiable by computing
the following model which is expressible in LIA: 

\smallskip
\noindent
D1.~~~\makebox[23mm][l]{${\tt new1(M,N)}$} \makebox[7mm][l]{$~\equiv~$} \makebox[20mm][l]{{\tt M=N}}

\noindent
D2.~~~\makebox[23mm][l]{${\tt new2(N)}$} \makebox[7mm][l]{$~\equiv~$} \makebox[20mm][l]{{\tt true}}

\noindent
D3.~~~\makebox[23mm][l]{${\tt diff(H,Na,N1)}$} \makebox[7mm][l]{$~\equiv~$} \makebox[20mm][l]{{\tt H+Na=N1}}

\smallskip
\noindent
Indeed, by replacing the left-hand side atoms by the corresponding right-hand side
LIA formulas in the final set of clauses~{\tt10},~{\tt11},~{\tt12f} 
and {\tt14}\,--{\tt18}, we get a set of valid 
implications.
Note that, by D3 we have that {\tt diff(H,Na,N1)} is a functional relation (that is, the usual `{\tt +}' on integers)
from {\tt (H,Na)} to {\tt N1}.

%

Thus, we have proved that Property {\it Sum} holds for the given 
program~{\it InsertionSort}.

\smallskip

\noindent
Let us conclude this section by commenting on the relationship between  
 difference predicates and lemmata.
If in clause~{\tt13} defining the difference predicate {\tt diff} we replace
its head {\tt diff(H,Na,N1)} by the constraint \mbox{\tt H\,+\,Na\,=\,N1}  
computed by the solver Z3,
we exactly get the CHC translation of Lemma~$L$ 
needed for proving
Property~{\it Sum} by structural induction 
on lists. Thus, the introduction of difference predicates 
can be viewed, at least in some cases, as a way of generating the lemmata required during proofs
by structural induction.

%% file: sections/35_rotate.tex
\label{rotate}
\newcommand{\eql}{$=_{\mathtt{list}}$}
\newcommand{\neql}{$\neq_{\mathtt{list}}$}

{In this section, we show through an example
that during the transformation Step~4 presented in 
Section~\ref{diffpred}, we can introduce,
instead of difference predicates, some auxiliary queries which correspond to 
lemmata required in the proof of the property of interest.}

Let us consider the following functional program {\it Rotate}, written in the
OCaml syntax, which defines: (i)~the familiar 
{\tt append} function which concatenates two lists, 
(ii)~the {\tt len} function which computes the length of a list, and
(iii)~the {\tt rotate} function which computes the circular rotation of 
a given list by~{$\mathtt {m~(\geq0)}$} positions. For instance, 
{\tt rotate 2 [7,4,5,9,1] = [5,9,1,7,4]}.

\begin{verbatim}
type list = Nil | Cons of int * list
let rec append l1 l2 =
   match l1 with
   | Nil -> l2
   | Cons(h,t) -> Cons(h, append t l2)
let rec len l =
   match l with
   | Nil -> 0
   | Cons(h,t) -> 1 + len t
let rec rotate m l = 
  if m<=0 then l else
    match l with
    | Nil -> Nil
    | Cons(h,t) -> rotate (m-1) (append t (Cons(h,Nil)))
\end{verbatim}

\noindent
Let us suppose that  we want to prove the following Property $\mathit{Rotation}$ stating that: 

\smallskip

$\mathtt{\forall\, l, k.~~  rotate ~(len~l)~(append~l~k)~=~append~k~l}$ \hfill 
\makebox[32mm][l]{$($Property~$\mathit{Rotation})$}

\smallskip

\noindent This property is used as a running example in a paper by 
Alan Bundy~\cite{Bun01} on the automation of proofs by mathematical induction. 
In that paper
the author discusses the issue of how to generate the lemmata needed 
for the inductive proofs and suggests the introduction of formulas with 
second order variables. 
Unfortunately, those second order variables require the use 
of second order unification and 
narrowing~(for these concepts the reader may refer to 
the paper by Baader and Snyder~\cite{BaS01}).

Now we will see how, with the help of the {\tt rotate} example, 
the difficulties due to the use of second order variables can be overcome and 
the required lemmata can be generated by applying
a variant of the technique proposed in Section~\ref{diffpred}. 
In particular, Property~{\it Rotation} is translated into a 
query (see clause~{\tt1} below) and
the lemmata needed for the satisfiability proof of that query 
{are introduced, not in the form of difference predicates, but in 
the form of auxiliary queries
(see Step~4* below)  whose satisfiability should in turn be proved.}

As in the example considered in Sections~\ref{chcsat} and~\ref{diffpred}, in the {\tt rotate}
example here we start off by
translating the initial functional program into a set of CHCs. By doing so we get:

\begingroup
\catcode`\@=\active
\def @{\neql}        

\catcode`\S=\active
\def S{$\leq$}       

\begin{verbatim}
 1. false :- len(L,M), append(L,K,W), rotate(M,W,Z), append(K,L,Z1), Z@Z1.
 2. append([],Ys,Ys). 
 3. append([H|Xs],Ys,[H|Zs]) :- append(Xs,Ys,Zs).
 4. len([],0).
 5. len([H|T],M) :- M=N+1, len(T,N).
 6. rotate(M,L,L) :- MS0.
 7. rotate(M,[],[]) :- M>0.
 8. rotate(M,[H|T],Z) :- M>0, N=M-1, append(T,[H],R), rotate(N,R,Z).
\end{verbatim}
\endgroup

\noindent 
where query {\tt 1} translates Property {\it Rotation}. For the 
clauses~{\tt2}\,--\,{\tt8} 
we have that {\tt append(Xs,Ys,Zs)}, {\tt len(L,M)}, and {\tt rotate(M,L,L1)} hold
iff $\mathtt{append~Xs~Ys}$ $=$ $\mathtt{Zs}$, 
$\mathtt{len~L}$ $=$ $\mathtt{M}$, and
$\mathtt{rotate~M~L}$ $=$ $\mathtt{L1}$, respectively, hold in the given 
functional program.

In what follows, we will 
use the predicates \eql~ and \neql~ to denote list equality 
and disequality, respectively.
Now, similarly to Section~\ref{diffpred}, we would like to transform 
clauses~{\tt 1}--\,{\tt 8} into an equisatisfiable set of clauses 
where no list variables occur.
In this transformation we apply the
Elimination Algorithm and we start off by introducing a new predicate 
{\tt new1} using the following clause:

\smallskip

\noindent
{\tt ~9.~new1(M) :- len(L,M), append(L,K,W), rotate(M,W,Z), append(K,L,Z1),\,Z\,\neql\,Z1.}

\smallskip

\noindent
where the argument~{\tt M} of {\tt new1} is the only integer variable in the body of query {\tt 1}. 
Then, we fold query {\tt 1} using clause {\tt 9}, and we get:

\smallskip
\noindent
{\tt 10.~false :- new1(M).}

\smallskip

We proceed by performing some unfolding steps starting from the {\tt len} predicate in clause~{\tt9}. 
After  suitable variable renamings we derive the following two clauses:

\smallskip

\noindent
{\tt 11.~new1(0) :- append(A,[],B), A\,\neql\,B.}

\noindent
{\tt 12.~new1(A) :- A=B+1, B$\geq$0, len(C,B), append(C,D,E), 
append(E,[F],G), }

\hspace{24.5mm}{\tt  rotate(B,G,H),   append(D,[F|C],I), H\,\neql\,I.}

\smallskip
\noindent
The list variables in clause {\tt 11} are eliminated by
introducing a new predicate {\tt new2}, defined by the clause:

\smallskip
\noindent
{\tt 13.~new2 :- append(A,[],B), A\,\neql\,B.}

\smallskip
\noindent
which is then used for folding clause  {\tt 11}. By folding, we get:

\smallskip
\noindent
{\tt 11f.~new1(0) :- new2.}

\smallskip
\noindent
The recursive definition of {\tt new2}, derived by an iteration of the
Elimination Algorithm, consists of the following clause only:

\smallskip
\noindent
{\tt 14.~new2 :- new2.}

\smallskip
\noindent
Now, in order to fold clause~{\tt12} using clause~{\tt9} and 
derive a recursive definition of {\tt new1}, 
we perform the following six steps, which are similar to those presented in 
Section~\ref{diffpred},
with the exception that, as already mentioned, we introduce auxiliary queries,
 instead of difference predicates.

\smallskip
\noindent
$\bullet$ {\it Step} 1. {\it Embed}. 
We have that the body of clause~{\tt9} is {embedded}
in the body of clause~{\tt12}, that is,  for each occurrence $A$ of an atom 
in the body of clause~{\tt9} there is an occurrence of an atom
in the body of clause~{\tt12} which is an instance of $A$. 

\smallskip
\noindent
$\bullet$ {\it Step} 2. {\it Rename}. We rename apart clause~{\tt 9}, 
which we would like to use for folding. We get:

\smallskip
\noindent
{\tt 9a.~new1(Ma) :- len(La,Ma), append(La,Ka,Wa), rotate(Ma,Wa,Za),}\nopagebreak

\hspace{27mm}{\tt append(Ka,La,Z1a), Za\,\neql\,Z1a.}                            

\smallskip
\noindent
$\bullet$ {\it Step} 3. {\it Match}. We match the body of clause~{\tt9a}  
against 
the body of clause~{\tt12} to be folded. We have that the conjunction \,{\tt len(La,Ma),\,rotate(Ma,Wa,Za)}\, in the body of clause {\tt9a} matches the
conjunction \linebreak\,{\tt len(C,B),\,rotate(B,G,H)}\, 
in the body of clause {\tt 12}
via the substitution 
$\sigma= \{{\mathtt{La/C,}}$ ${\mathtt{Ma/B,}}$ 
${\mathtt{Wa/G,}}$ ${\mathtt{Za/H}}\}$. 
By applying the substitution~$\sigma$ to clause~{\tt9a}, we get:\nopagebreak

\smallskip
\noindent
{\tt 9m.~new1(B) :- len(C,B), append(C,Ka,G), rotate(B,G,H),}\nopagebreak
	
\hspace{25mm}{\tt append(Ka,C,Z1a), H\,\neql\,Z1a.}                            

\smallskip
\noindent
Thus, the mismatching conjunction of clause {\tt 9m} is:\nopagebreak

\smallskip
\noindent
{\tt ~M.~append(C,Ka,G), append(Ka,C,Z1a), H\,\neql\,Z1a}.  

\smallskip
\noindent
and the mismatching conjunction of clause {\tt 12} is:

\smallskip
\noindent
{\tt ~N.~append(C,D,E), append(E,[F],G), append(D,[F|C],I), H\,\neql\,I}.
	
\smallskip
\noindent
$\bullet$ {\it Step} 4*. {\it Introduce Auxiliary Queries}. 
Now, in order to fold 
clause~{\tt12} using clause~{\tt 9m}, we need to replace 
the mismatching conjunction {\tt N} of clause~{\tt12} 
by the mismatching conjunction {\tt M} of clause~{\tt9m}. 
This replacement cannot be done by applying the technique
presented in 
Section~\ref{diffpred}, where we have also 
introduced the difference predicate {\tt diff}.
Indeed, no output integer variables occur in the 
mismatching conjunctions and the associated {\tt diff} 
predicate would have no arguments at all.

Thus, we will follow an alternative path:
(i)~first, we will do the replacement of {\tt N} by {\tt M} (see Step~$5$ below),
and then (ii)~we will prove, as an auxiliary lemma,
the soundness of that replacement.
As indicated in Section~\ref{diffpred} (see Property {\tt I}), this auxiliary lemma, 
call it {\tt L1}, is
$\mathtt{\forall (N \rightarrow \exists 
\overline{Y}.\,M)}$, where
$\mathtt{\overline{Y}}$ is the tuple of the variables occurring in {\tt M} 
and not in the rest of clause {\tt 9m}.
By using the definitions of the conjunctions {\tt N} and {\tt M}, 
Lemma~{\tt L1} can be written as follows:




\smallskip
\noindent
{\tt  L1.~} $\forall$ {\tt  (append(C,D,E), append(E,[F],G), append(D,[F|C],I), 
H\,\neql\,I} $\mathtt{\rightarrow}$\nopagebreak

\hspace{21mm}$\exists$\,{\tt Ka,Z1a.\,append(C,Ka,G), append(Ka,C,Z1a), 
H\,\neql\,Z1a)}

\smallskip

%
%
%
\noindent
This universally quantified implication {\tt  L1} is not in CHC form, 
and in order 
to prove it by using our transformation technique, we first need to transform it
into a set of CHCs as indicated in the following 
Steps~(4*.1)\,--\,(4*.4).

\smallskip
\noindent

	
\smallskip
\noindent
{\it Step} 4*.1. {\it Move conclusion to premise}.  We move the conclusion of 
{\tt L1} to the premise and we get a new universally quantified implication:
\nopagebreak

\smallskip

\noindent
{\tt H1.}~$\forall$ {\tt  (append(C,D,E), append(E,[F],G), append(D,[F|C],I), 
H\,\neql\,I,} 

\hspace{13mm}{\tt (}$\neg \exists$\,{\tt Ka,Z1a.~append(C,Ka,G), append(Ka,C,Z1a), 
H\,\neql\,Z1a)} $\mathtt{\rightarrow}$ {\tt false)}

\smallskip
\smallskip
\noindent
{\it Step} 4*.2. {\it Use Functionality and Totality of\, {\tt append}, 
and Properties of~\eql}. 
Since {\tt append(Xs,Ys,Zs)} denotes a total functional relation from 
{\tt (Xs,Ys)} to {\tt Zs}, from {\tt H1} 
we get that the following universally quantified equivalence holds:

\smallskip

$\forall$ {\tt ((}$\neg \exists$\,{\tt Ka,Z1a.~append(C,Ka,G), append(Ka,C,Z1a), H\,\neql\,Z1a)} ~~$\leftrightarrow$

\noindent
\hspace{11mm}{\tt (}$\neg \exists$\,{\tt Ka.~append(C,Ka,G)} $\vee$ 

\noindent
\hspace{11mm}{\tt (}\hspace{2.7mm}$\exists$\,{\tt Ka,Z1a.~append(C,Ka,G), append(Ka,C,Z1a), H\,\eql\,Z1a))}

\smallskip
\noindent
Thus, by using the distributive law, we rewrite {\tt H1}
into the conjunction of the following two universally quantified implications:

\smallskip

\noindent
{\tt H2.}~$\forall$ {\tt  (append(C,D,E), append(E,[F],G), append(D,[F|C],I), H\,\neql\,I,}\nopagebreak

~~~~~~{\tt (}$\neg \exists$\,{\tt Ka.~append(C,Ka,G))} $\mathtt{\rightarrow}$ {\tt false)}

\smallskip

\noindent
{\tt H3.}~$\forall$ {\tt  (append(C,D,E), append(E,[F],G), append(D,[F|C],I), H\,\neql\,I,}\nopagebreak

~~~~~~{\tt (}$\exists$\,{\tt Ka,Z1a.~append(C,Ka,G), append(Ka,C,Z1a), H\,\eql\,Z1a)} $\mathtt{\rightarrow}$ {\tt false)}

\smallskip
\noindent
Now, since {\tt H} has a single occurrence in the premise of {\tt H2} and 
$\mathtt{\forall\, I.~\exists\, H.~H \mbox{\,\neql\,} I}$,
we can remove {\tt H\,\neql\,I} from {\tt H2}. Then, {\tt I} has a single occurrence in the 
implication derived after removal and, by the totality of  {\tt append}, we 
can  remove also
{\tt append(D,[F|C],I)}. Thus, from {\tt H2} we get:

\smallskip

\noindent
{\tt H4.}~~$\forall$ {\tt  (append(C,D,E), append(E,[F],G), (}$\neg \exists$\,{\tt Ka.~append(C,Ka,G))} $\mathtt{\rightarrow}$ {\tt false)}

\smallskip
\smallskip
\noindent
{\it Step} 4*.3. {\it Derive CHC Queries}.  (i)~First we replace $\neg \exists$\,{\tt Ka.\,append(C,Ka,G)}
in {\tt H4} by a new predicate 
{\tt not\_exists\_2nd\_append(C,G)}, whose defining clauses 
will be derived at the following
Step~4*.4, and then (ii)~we remove the existential 
quantification from the premise of~{\tt H3} (this removal preserves equivalence).
By doing so, we get the following two CHC queries:

\smallskip
\noindent
{\tt Q1.1 ~false :- append(C,D,E), append(E,[F],G), not\_exists\_2nd\_append(C,G).}

\noindent
{\tt  Q1.2 ~false :- append(C,D,E), append(E,[F],G), append(D,[F|C],I), 
H\,\neql\,I,}

\hspace{24.5mm}{\tt append(C,Ka,G), append(Ka,C,Z1a), H\,\eql\,Z1a.}


\smallskip
\noindent
{\it Step} 4*.4. {\it Eliminate Negation}. We derive CHCs defining predicate
{\tt not\_exists\_2nd\_append(Xs,Ys)}
by using well-known techniques for eliminating negation from
logic programs (see, for instance, the {\em Negation Technique}~\cite{SaT84} and 
the {\it UFS} transformation strategy~\cite{PeP00a}):


\begingroup
\catcode`\D=\active
\def D{$\neq$}       

\begin{verbatim}
15. not_exists_2nd_append([X|Xs],[]).
16. not_exists_2nd_append([X|Xs],[Y|Ys]) :- XDY.
17. not_exists_2nd_append([X|Xs],[Y|Ys]) :- X=Y, not_exists_2nd_append(Xs,Ys).
\end{verbatim}
\endgroup

\smallskip
\noindent
$\bullet$ {\it Step} 5. {\it Replace}. Now by applying Lemma {\tt L1},
we replace, in clause~{\tt 12}, conjunction~{\tt N}
by conjunction~{\tt M}, and we get the following clause:

\smallskip
\noindent
{\tt 12r.~new1(A) :- A=B+1, B$\geq$0, len(C,B), append(C,Ka,G), rotate(B,G,H),
	
\hspace{26mm}append(Ka,C,Z1a), H\,\neql\,Z1a.}

\smallskip

\noindent
$\bullet$ {\it Step} 6. {\it Fold}. We fold clause~{\tt12r} using clause~{\tt9m},
and we get:\nopagebreak

\smallskip
\noindent
{\tt 12f.~new1(A) :- A=B+1, B$\geq$0, new1(B).}

\medskip
\noindent
The clauses derived after Steps 1--\,6 are: {\tt 10}, {\tt 11f}, {\tt 14}, 
{\tt 12f}, {\tt 15}, {\tt 16}, {\tt 17}, together with 
queries {\tt Q1.1} and {\tt Q1.2}, and clauses~{\tt 2}\,--\,{\tt 8} belonging to 
the initial set.

\smallskip
Now, we are left with the task of proving Lemma {\tt L1}.
Since Steps~(4*.1)\,--\,(4*.4) preserve satisfiability,
this proof  can be done by showing the satisfiability of the two 
queries {\tt Q1.1} and {\tt Q1.2}.
Our transformation continues starting from those two queries
by following a strategy similar to the one we have 
applied above starting from the
initial query~{\tt 1}. 
Thus, we introduce the following two new definitions:

\smallskip
\noindent
{\tt 18.~new3(F) :- append(C,D,E), append(E,[F],G), not\_exists\_2nd\_append(C,G).}

\noindent
{\tt  19.~new4(F) :- append(C,D,E), append(E,[F],G), append(D,[F|C],I), 
H\,\neql\,I,}

\hspace{24.5mm}{\tt append(C,Ka,G), append(Ka,C,Z1a), H\,\eql\,Z1a.}

\smallskip
\noindent
Then, we fold  {\tt Q1.1} and {\tt Q1.2} using those clauses, and we get:

\smallskip
\noindent
{\tt 20.~false :- new3(F).}

\noindent
{\tt  21.~false :- new4(F).}

\smallskip
\noindent
After some more transformation steps, we derive the following
 final set of clauses, called~{\it TransfCls}: 

\medskip
\begingroup
\catcode`\D=\active
\def D{$\neq$}       

\catcode`\G=\active
\def G{$\geq$}       

\noindent
\hspace*{-3mm}
\begin{tabular}{l@{\hspace{16mm}}l}
\tt10.  false :- new1(A).      &  \\
\tt11f.~new1(0) :- new2.       & \tt26.  new6(A,B) :- new7(B)  \\
\tt12f.~new1(A) :- A=B+1, BG0, new1(B).   & \tt27.  new6(A,B) :- BDC, new8(C).  \\
\tt14.  new2 :- new2.          & \tt28.  new6(A,B) :- new8(B).  \\
\tt20.  false :- new3(A).      & \tt29.  new6(A,B) :- new6(A,B).  \\
\tt21.  false :- new4(A).      & \tt30.  new7(A) :- BDA, new8(B).  \\
\tt22.  new3(A) :- new3(A).    & \tt31.  new7(A) :- new8(A).  \\
\tt23.  new4(A) :- new5(A).    & \tt32.  new8(A) :- new8(B).  \\
\tt24.  new5(A) :- new5(A).    & \tt33.  false :- new9(A). \\ 
\tt25.  false :- new6(A,F).    & \tt34.  new9(A) :- new9(A).
\end{tabular}

\endgroup

\medskip
\noindent
During the derivation of the above clauses we generate the following two extra Lemmata
{\tt L2} and {\tt L3} and we introduce the corresponding two auxiliary queries
{\tt Q2} and {\tt Q3}, respectively:

\smallskip
\noindent
{\tt L2.} $\forall$ {\tt (append(C,[A,F|B],G), append(I,[F|B],H), G\,\neql\,H} ~$\rightarrow$\nopagebreak

\hspace{20.5mm}{\tt (}$\exists$  {\tt F1,G1.~append(C,[A|B],F1),~append(I,B,G1),~F1\,\neql\,G1))}

\smallskip
\noindent
{\tt Q2.~false :- append(C,[A,F|B],G), append(I,[F|B],H), G\,\neql\,H,}\nopagebreak

\hspace{20.5mm}{\tt append(C,[A|B],F1),~append(I,B,G1),~F1\,\eql\,G1.}

\medskip
\noindent
{\tt L3.} $\forall$ {\tt (append(B,[A|C],D)}  ~$\rightarrow$  $\exists$ {\tt B1.append(B1,C,D))}\nopagebreak

\smallskip
\noindent
{\tt Q3.~false :- append(B,[A|C],D), {not\_exists\_1st\_append(C,D).}}

\smallskip
\noindent
where {\tt not\_exists\_1st\_append(C,D)} is a predicate equivalent to 
$\neg\exists$ {\tt B1.append(B1,C,D)}.

The final set {\it TransfCls} of clauses
has no list arguments and all constraints are LIA formulas.
As in the example of Section~\ref{diffpred},
it can be shown that the transformation steps we have performed,
are sound, and thus if we are able to prove the satisfiability of the clauses 
of {\it TransfCls}, then Property~{\it Rotation} holds.

Now, the CHC solver Z3 easily proves that the set {\it TransfCls} of clauses is satisfiable.
Indeed, every clause in {\it TransfCls} has one atom in its body, and hence
the interpretation that assigns the truth value {\tt false} to every
predicate is a model.
In particular, the satisfiability of {\it TransfCls}
shows also the validity of the various lemmata we have generated
during the derivation.

This concludes the proof that Property {\it Rotation} holds for the given 
{\tt rotate} function.

%% file: sections/4_conclremks.tex
\label{conclrems}
Let us briefly discuss on the soundness of the transformation technique
presented in this paper through a couple of examples 
in Sections~\ref{diffpred} and~\ref{rotate} and
also on the mechanization of that technique.


The notion of soundness we have used is defined by the following
property: if the clauses obtained after transformation are 
satisfiable, then so are the clauses before transformation.
Thus, the satisfiability of the clauses obtained after transformation
is sufficient to guarantee that the property of the 
functional program that we want to verify indeed holds.

The crucial hypothesis needed to show the soundness of our
transformation technique is that the predicates occurring in the
initial set of clauses define total functional relations.
This property is enforced by construction, whenever those predicates 
are the CHC translation of functions that terminate for all inputs. 
Moreover, to show soundness,
we also use the fact that every lemma generated during 
derivation is an implication, and we replace, in the body of a clause,
an instance of the premise of the lemma by an instance of its conclusion.
If the lemmata are not equivalences, the transformations do not
necessarily derive final clauses that are equisatisfiable 
with respect to the initial
ones. However, as mentioned at the end of Section~\ref{diffpred},
equisatisfiability is guaranteed if {every} generated lemma corresponds to the
definition of a {\em functional} difference predicate. 
This functionality property can be checked 
in the model computed by the CHC solver, which is expressed as
a set of LIA and Bool constraints.

Concerning the mechanization of our transformation technique,
we need to extend the Elimination Algorithm~\cite{De&18a} with a suitable
automated mechanism for introducing difference predicates and/or
auxiliary queries.
As shown in Sections~\ref{diffpred} and~\ref{rotate}, 
this mechanism can be based on the result of matching
the clauses obtained by unfolding (see clause~{\tt12} in the 
{\it InsertionSort} example, and clause~{\tt12} in the {\it Rotate} example) 
against the predicate definitions
introduced in previous transformation steps (see clause~{\tt9} in both examples).
More sophisticated mechanisms may take into account the constraints
occurring in the clauses, and may apply widening techniques 
which have been 
considered in other transformation methods~\cite{De&14c,Ka&16}.
We have made initial steps towards an implementation of such an extended 
Elimination Algorithm using the VeriMAP transformation and 
verification system~\cite{De&14b}.

In order to evaluate the generality of our verification approach 
based on the Elimination Algorithm extended with lemma generation,
we have also done some experiments on various sorting algorithms
and we have semi-automatically proved various properties for a few of them~\cite{De&19b}.

To summarize, 
this paper presents ongoing work which follows a very general approach
to program verification based on constrained Horn clauses.
As shown in the examples we have presented, the reduction of a program
verification problem to a CHC satisfiability problem can often be 
obtained by a straightforward translation. However, proving the satisfiability
of the clauses obtained by that translation is, in many cases, a much harder
task. In a series of papers~\cite{De&14c,De&15c,De&17b,De&18c,De&18a,Ka&16,MoF17} 
it has been shown that by combining various 
transformation techniques, such as {\em Specialization} and 
{\em Predicate Pairing},
we can derive equisatisfiable sets of clauses where the efficacy of the
CHC solvers is significantly improved. This approach avoids the 
burden of implementing very sophisticated solving strategies depending 
on the class of satisfiability problems to be solved. 
In particular, in the class of problems considered in this paper
consisting in checking the satisfiability of clauses over inductively 
defined data structures, 
we can avoid to implement {\it ad hoc} strategies that deal with induction proofs.
We leave it for future work to experiment {on various
benchmarks available from the literature and to test whether the 
transformation-based approach pays off in practice.}

%% file: sections/5_acknowledgement.tex
\label{acknow}
All authors are members of the INdAM-GNCS Italian Research Group. 
Many thanks to the anonymous referees for their helpful suggestions
and constructive comments. 